\documentclass[aps,prl,twocolumn,superscriptaddress,10pt]{revtex4-2}
\usepackage{babel} 
\usepackage{amsmath}
\usepackage{amssymb}
\usepackage{amsfonts}
\usepackage{bm}

\usepackage{graphicx}

\usepackage{xcolor}

\usepackage[colorlinks=true,linkcolor=blue,urlcolor=blue,citecolor=blue,anchorcolor=blue]{hyperref}

\usepackage{braket}      
\usepackage{mathrsfs}    
\usepackage{bbm}         

\usepackage{comment}     
\usepackage[normalem]{ulem} 

\definecolor{npurple}{rgb}{0.6,0.3,0.8}

\begin{document}
\title{Temporal modulation as a resource: enhanced frequency estimation in continuous variable systems}

\author{Ningxin Kong}
\affiliation{State Key Laboratory of Artificial Microstructure and Mesoscopic Physics, School of Physics, Frontiers Science Center for Nano-optoelectronics, $\&$ Collaborative Innovation Center of Quantum Matter, Peking University, Beijing 100871, China}

\author{Qiongyi He}
\email{qiongyihe@pku.edu.cn}
\affiliation{State Key Laboratory of Artificial Microstructure and Mesoscopic Physics, School of Physics, Frontiers Science Center for Nano-optoelectronics, $\&$ Collaborative Innovation Center of Quantum Matter, Peking University, Beijing 100871, China}
 \affiliation{Hefei National Laboratory, Hefei 230088, China}
\affiliation{\mbox{Collaborative Innovation Center of Extreme Optics, Shanxi University, Taiyuan, Shanxi 030006, China}}

\author{Matteo G. A. Paris}
\email{matteo.paris@fisica.unimi.it}
\affiliation{Dipartimento di Fisica, Universit\`a di Milano, I-20132 Milano, Italy}

\begin{abstract} 
    Frequency estimation, a cornerstone of quantum metrology, has been significantly enhanced by advanced quantum sensing strategies. 
    However, most protocols rely either on static or time-independent encoding mechanisms, inherently limiting their achievable precision scaling,  or on control strategies requiring changing the Hamiltonian and/or implementing feedback mechanisms. To overcome this, we investigate a simpler dynamical encoding protocol where the quantum oscillator is driven by a general continuous temporal frequency modulation $\Omega(t) = \omega_0 f(t)$. We analytically demonstrate that for a given modulation profile $f(t)$ and its corresponding time-integral $F(t)$, the quantum Fisher information (QFI) scales as $\mathcal{O}(F(t)^2)$. This enhancement stems from the fact that temporal encoding fundamentally alters the mechanism of dynamical phase accumulation. Crucially, when evaluated under the energy and evolution-time constraints, this framework reveals a genuine precision enhancement over the conventional time-independent baseline. By analyzing explicit polynomial and exponential modulations, we establish that arbitrary precision scaling can be deterministically engineered, with ultimate bounds that are asymptotically saturable via optimal homodyne detection. Our framework provides a universal paradigm for exploiting time-dependent quantum control in next-generation sensors.
\end{abstract}
  
    \maketitle

\textit{Introduction.---}Accurate frequency estimation is central to quantum metrology~\cite{Hansch2002,Rafal2016,Cappellaro2017,Sloyd2018,Silberhorn2018,Felicetti2020,Gessner2025,Matteo2025a} and underlies a wide range of quantum technologies~\cite{Aharonov1999,Thew2007,chen2021}.
While quantum resources can enhance sensitivities beyond classical limits~\cite{Maccone2004,Maccone2006,Maccone2011,Mitchell2011,Zoller2024,Zelevinsky2024,Gao2025,Scarlino2025}, conventional protocols based on static state encoding remain fundamentally limited in precision scaling due to the inherent dependence of the system's eigenvalues and eigenvectors on the target frequency. Although free evolution under a time-independent Hamiltonian continuously imprints frequency information onto the quantum state, it remains an open question whether temporal control can fundamentally reprogram the encoding mechanism to achieve superior scaling, particularly when accounting for the total resource overhead of energy and sensing time.

Given the ubiquity of harmonic dynamics across physical platforms, the quantum harmonic oscillator (QHO) serves as a universal model for quantum metrology~\cite{Caves2011,Wilson2019,Braun2020,Biercuk2021,Leibfried2021,Smerzi2019,Kippenberg2019}. In such harmonic systems, the oscillator frequency governs both the static properties and dynamical evolution, making its precise estimation a central objective that has driven extensive theoretical and experimental efforts within the QHO framework. While recent studies have demonstrated that non-equilibrium control protocols, such as sudden frequency jumps, can successfully generate squeezing to boost sensitivity, these analysis typically restricted to highly specific control schemes~\cite{Matteo2025b,Zaidi1988,Rhodoes1989,Rego2020}. A universal understanding of how arbitrary continuous temporal modulation dictates parameter encoding and ultimate precision is still lacking.
This gap motivates two fundamental questions: (i) Can tailored temporal modulation as the resource systematically transcend traditional precision limits? (ii) Do such dynamic strategies yield a genuine metrological advantage when subjected to strict constraints on energy and evolution time?

In this Letter, we address all the above issues. Regarding the first question, we investigate a dynamic encoding scheme by driving the QHO with a general temporal frequency modulation $\Omega(t) = \omega_0 f(t)$, where the modulation profile $f(t)$ is independent of the unknown base frequency $\omega_0$, scaling as $\mathcal{O}(t^n)$. By quantifying the precision limit via the QFI, we reveal a universal QFI $\mathcal{F}_t(t)$ scaling as $\mathcal{O}(F(t)^2)$, with $F(t)$ denoting the time-integral of $f(t)$. We demonstrate that this significant enhancement fundamentally stems from altering the mechanism of dynamical phase accumulation during the time-dependent encoding process.
To address the second question, we define the QFI enhancement ratio that quantifies the genuine metrological gain relative to the time-independent encoding protocol. Subject to identical energy and sensing time constraints, our temporal modulation framework preserves a genuine gain scaling as $\mathcal{O}(F(t)^2 / [f(t)t^2])$.
To demonstrate the versatility of our framework, we analyze linear, quadratic, and exponential modulation profiles. In particular, the exponential driving envelope yields an asymptotic QFI scaling of $\mathcal{O}(e^{2t})$, which directly translates to a genuine enhancement of $\mathcal{O}(e^t t^{-2})$. Furthermore, this ultimate precision bound can be asymptotically saturated utilizing real-time optimized homodyne detection. Overall, our framework establishes temporal modulation as a metrological resource, offering a universally applicable paradigm for the optimal design of advanced continuous-variable quantum sensors.

\textit{Harmonic oscillator with frequency modulation.---}We consider a QHO subject to a time-dependent frequency modulation $\Omega(t)$. Setting $\hbar=1$, the system dynamics are governed by the Hamiltonian 
\begin{equation}
    \hat{H}(t)=\frac{1}{2}\left[ \hat p^2+\Omega^2(t) \hat x^2\right],
    \label{eq1}
\end{equation}
where $\hat x$ and $\hat p$ are the canonically conjugate position and momentum operators, satisfying $[\hat x, \hat p]=i$. 
The time-dependent Schr\"odinger equation, $i\partial_t \Psi(x,t)=\hat H(t)\Psi(x,t)$, can be mapped onto an equivalent time-independent problem to determine the state at any arbitrary time $t > 0$~\cite{Seleznyova1995}. Because the Hamiltonian is strictly quadratic, an initially Gaussian state remains Gaussian throughout the evolution. Consequently, we employ a generalized Gaussian ansatz for the wave function in coordinate representation,
\begin{equation}
    \Psi(x,t)=\frac{1}{\sqrt{r(t)}}e^{i\mu(t) \hat{x}^2}\,\phi(\xi,\tau(t)),
    \label{eq2}
\end{equation}
where $\xi=\hat{x}/r(t)$ represents a squeezing transformation, $\mu(t)$ acts as a shearing parameter, and $\tau(t)$ is a rescaled time coordinate. Substituting Eq.~(\ref{eq2}) into the Schr\"odinger equation reveals that the effective unitary evolution operator takes the following form (See details in Supplemental Material (SM)~\cite{sm})
\begin{equation}
    \hat U_z(t) 
    = e^{i \frac{\mu(t)}{2\omega_0}(\hat{a}+\hat{a}^\dagger)^2}
    \, e^{-\frac{\theta(t)}{2}(\hat{a}^{\dagger 2}-\hat{a}^2)}
    \, e^{-i \gamma(t)\left(\hat{a}^\dagger \hat{a} + \frac{1}{2}\right)},
    \label{eq3}
\end{equation}
where the bosonic operators $\hat a$ and $\hat a^\dagger$ are defined with respect to a fixed reference frequency $\omega_0$ through $\hat{x} = (\hat{a} + \hat{a}^\dagger)/\sqrt{2\omega_0}$ and $\hat{p} = i\omega_0(\hat{a}^\dagger - \hat{a})/\sqrt{2 \omega_0}$. This canonical transformation maps the phase-space operators ($\hat x$,$\hat p$) onto a fixed bosonic mode basis. The time-dependent shearing $\mu(t)$, squeezing $\theta(t)$, and rotation $\gamma(t)$ parameters are entirely determined by the modulation $f(t)$ via a complex auxiliary function $z(t)$
\begin{equation}
    \begin{aligned}
        r(t) & = |z(t)|, \quad \mu(t) = \frac{\dot r(t)}{2r(t)},\\
        \theta(t) & = \ln r(t),\quad \gamma(t) = \arg z(t)\,.
    \end{aligned}
    \label{eq4}
\end{equation}
Crucially, the dynamic of $z(t)$ is governed by the classical equation of motion for the oscillator,
$\ddot z(t) + \Omega^2(t)z(t) = 0$ subject to the initial conditions $z(0)=1$ and $\dot z(0) = i \omega_0$.

\textit{Temporal modulation protocol.---}In quantum metrology, the precision of parameter estimation under a unitary encoding $\hat{U}_g$ is fundamentally dictated by the distinguishability between neighboring output states, $|\Psi_g\rangle = \hat{U}_g|\Psi\rangle$ and $|\Psi_{g+\delta g}\rangle = \hat{U}_{g + \delta g}|\Psi\rangle$. This distinguishability is quantified by the QFI, $\mathcal{F}_g = 4\text{Var}(\hat{h})_{|\Psi\rangle}$, where $\hat{h} = i\hat{U}_g^\dagger \partial_g \hat{U}_g$ defines the local Hermitian generator associated with $g$, and $\text{Var}(\cdot)$ is the variance with the initial state $|\Psi\rangle$. The QFI establishes the ultimate lower bound on the root-mean-square error $\Delta g$ via the quantum Cramér-Rao bound (QCRB), $\Delta g = 1/\sqrt{\nu \mathcal{F}_g}$, where $\nu$ is the number of independent measurements. This bound is saturable in the asymptotic limit $\nu \to \infty$ given an optimal measurement and estimator strategy~\cite{Caves1994,Augusto2009,Kok2010}. 

Generally, we consider a QHO with a temporal frequency modulation $\Omega(t) = \omega_0 f(t)$, where $\omega_0$ is the parameter of interest and 
$f(t) = 1 + \sum_{k=1}^n a_k t^k$ is an arbitrary $n$-th order polynomial independent of $\omega_0$. We focus on the adiabatic regime, characterized by a slowly varying frequency satisfying $|\dot\Omega(t)/\Omega(t)^2 | \ll 1$. 
This condition is naturally satisfied in the high-frequency limit, $\omega_0 \gg 1$, which is routinely accessed in a wide range of experimental platforms, including atomic systems, microwave circuits, and mechanical resonators~\cite{Lan2021,Clerk2020,Sillanp2018}. Furthermore, adiabatic parameter variations are typically easier to implement experimentally than abrupt quenches. The present analysis, therefore, applies directly to experimentally relevant operating conditions.
Under this approximation, the general solution for $z(t)$ takes the standard Wentzel-Kramers-Brillouin (WKB) form
\begin{equation}
    z(t)=\frac{1}{\sqrt{\Omega(t)}}\left(Ae^{i\Phi(t)}+Be^{-i\Phi(t)}\right),
    \label{eq5}
\end{equation}
where $A=\sqrt{\omega_0}\left(1-\frac{ia_1}{4\omega_0}\right)$ and
$B=\frac{ia_1}{4\sqrt{\omega_0}}$ are determined by the initial conditions, and the accumulated phase is defined by $\Phi(t)=\int_0^t \Omega(s)\,ds =\omega_0 F(t)$. The integrated modulation profile, $F(t) = \int_0^t f(s) ds$, admits a power-series expansion $F(t) = \sum_{k=1}^{n+1} \frac{a_{k-1}}{k} t^{k} $ with $a_0 = 1$.

By substituting the WKB solution for $z(t)$ into Eq.~(\ref{eq4}), we obtain the explicit time evolution of the dynamical parameters $\mu(t)$, $\theta(t)$, and $\gamma(t)$. To evaluate the metrological sensitivity to the base frequency $\omega_0$, we extract the derivatives of these parameters with respect to $\omega_0$. Thus, in the adiabatic regime and in the long-time limit, the QFI for estimating $\omega_0$ from the evolved state $\ket{\Psi}_{\omega_0} = \hat{U}_z(t) \ket{\Psi}$ is entirely determined by the local generator
\begin{equation}
    \hat{h} = i\, \hat U_z^\dagger(t)\, \partial_{\omega_0}\hat U_z(t) = - P_\mu \hat{h}_\mu + P_\theta \hat{h}_\theta + P_\gamma \hat{h}_\gamma,
    \label{eq7}
\end{equation}
where the sensitivities of the respective dynamical parameters as
\begin{equation}
    \begin{aligned}
        P_\mu & = \partial_{\omega_0}\left[\frac{\mu(t)}{2\omega_0}\right] \simeq -\frac{a_1 f(t)F(t)\sin2\Phi(t)}{4\omega_0 D^2},\\
        P_\theta & = \partial_{\omega_0}\theta(t) \simeq \frac{F(t)(2\lambda \cos 2\Phi(t) + \lambda^2 \sin 2\Phi(t))}
        {2 D},\\
        P_\gamma & = \partial_{\omega_0}\gamma(t) \simeq \frac{F(t) }{D}.
    \end{aligned}
\end{equation}
where $ \lambda=a_1/(2\omega_0) \ll 1$ is a dimensionless perturbative parameter, and $D = 1 + \lambda \sin 2\Phi(t) + \lambda^2 \sin^2\Phi(t)$ remains bounded for all $t$, implying $D \sim \mathcal{O}(1)$.
The corresponding constituent operators are defined as $\hat{h}_\mu = e^{2\theta(t)} ( 1 + 2 \hat{a}^\dagger \hat{a} + e^{2i\gamma(t)} \hat{a}^{\dagger 2} + e^{-2i\gamma(t)} \hat{a}^2 )$, $\hat{h}_\theta = i ( e^{2i\gamma(t)} \hat{a}^{\dagger 2} - e^{-2i\gamma(t)} \hat{a}^2 )$, and $\hat{h}_\gamma = \hat{a}^\dagger \hat{a} + 1/2$, with $e^{2\theta(t)}=D/f(t)$.
Although the sensitivities of the shearing [$P_\mu$] and squeezing [$P_\theta$] parameters are formally proportional to the diverging terms $f(t)F(t)$ and $F(t)$, respectively, their growth is heavily suppressed by oscillating factors and the adiabatic condition. 
In contrast, the sensitivity of the rotation parameter $P_\gamma$ scales directly as $F(t)/D$, indicating that the overall sensitivity to $\omega_0$ is dominated by the dynamical phase $\gamma(t)$.

Consequently, the total QFI $\mathcal{F}_t(t) = 4\, \mathrm{Var}(\hat h)$, naturally decomposes into four distinct contributions: the shearing [$\mathcal{F}_{\mu}$], the squeezing [$\mathcal{F}_{\theta}$], the rotation [$\mathcal{F}_{\gamma}$] terms, alongside a cross-correlation interference term [$\mathcal{F}_c$] which can be negative (destructive) or positive (constructive) depending on evolution
\begin{equation}
    \begin{aligned}
        \mathcal{F}_{\mu} & \simeq \frac{a_1^2F(t)^2\sin^2 2\Phi(t)}{4\omega_0^2D^2}
    \Big( 8\text{Re}[e^{-2i\gamma(t)}\alpha^2] + 8|\alpha|^2+2\Big), \\
        \mathcal{F}_{\theta} & \simeq \frac{F(t)^2 (2\lambda \cos 2\Phi(t) + \lambda^2 \sin 2\Phi(t))^2  (4|\alpha|^2 + 2)}{ D^2},\\
        \mathcal{F}_{\gamma} & \simeq \frac{4 F(t)^2 |\alpha|^2}{D^2},\\
        \mathcal{F}_c & = -4 P_\mu P_\theta \text{Cov}(\hat h_\mu,\hat h_\theta) 
        - 4 P_\mu P_\gamma \text{Cov}(\hat h_\mu,\hat h_\gamma) \\
        & + 4 P_\theta P_\gamma \text{Cov}(\hat h_\theta,\hat h_\gamma),
    \end{aligned}
    \label{eq9}
\end{equation}
where the symmetric covariance $\text{Cov}(\hat{A},\hat{B}) = \langle \hat{A}\hat{B} + \hat{B}\hat{A} \rangle - 2 \langle \hat{A}\rangle \langle \hat{B} \rangle$ for the initial state $\ket{\Psi}$. 
Within the adiabatic regime, the shearing and squeezing contributions are strongly suppressed by rapid oscillations and the adiabatic constraint. As a result, the rotation-induced component, $\mathcal{F}_{\gamma}$, directly driven by the sustained phase accumulation $F(t)^2$, overwhelmingly dominates total QFI. Ultimately, the QFI exhibits asymptotic scaling $\mathcal{F}_t(t) \sim \mathcal{O}(t^{2(n+1)})$ for $f(t) \sim \mathcal{O}(t^n)$ and $\mathcal{F}_t(t) \sim \mathcal{O}(e^{2t})$ for $f(t) \sim \mathcal{O}(e^t)$, demonstrating that the precision enhancement in estimating $\omega_0$ via temporal frequency modulation originates primarily from reshaping the phase accumulation process, which can be deterministically tailored through the modulation function.

\textit{Energy constraints and genuine metrological gain.---}To rigorously assess the genuine metrological advantage provided by the temporal modulation under fair resource constraints, we evaluate the QFI enhancement ratio $\mathcal{R}(t)$, while fixing both the available energy and the sensing time. Here, $\mathcal{F}_t(t)$ and $\mathcal{F}_0(t)$ represent the QFIs for estimating $\omega_0$ in the driven and un-driven QHO schemes, respectively. We assume the system is initialized in a coherent state, denoted by $|\alpha\rangle$ for the modulated scheme and $|\alpha_0\rangle$ for the time-independent reference. For the standard un-driven protocol ($\Omega = \omega_0$), the QFI reduces to $\mathcal{F}_0(t) = 4 t^2 |\alpha_0|^2$, with the energy given by $E_0 = \omega_0 (|\alpha_0|^2+1/2)$. Conversely, for the modulated scheme with $\Omega(t) = \omega_0 f(t)$, the instantaneous energy expectation value, $E(t) = \langle \alpha | \hat{H}(t) | \alpha \rangle$, can be explicitly decomposed as:
\begin{equation}
    \begin{aligned}
    E(t) & \simeq \Omega(t)
        \Big[
        |\alpha|^2 + \frac{1}{2}
        + \frac{a_1}{\omega_0}\alpha_1\alpha_2
        + \frac{a_1^2}{4\omega_0^2}\alpha_1^2
        + \frac{a_1^2}{16\omega_0^2}
        \Big], \\
    \end{aligned}   
    \label{eq10}
\end{equation}
which scales as $f(t)$.
To ensure a fair comparison, we impose an energy-constraint condition by matching the mean energy of the time-independent to the final instantaneous energy of the driven system, $E_0 = E(t)$. This determines the effective resource allocation required for the static probe. Under this constraint, the QFI enhancement ratio $\mathcal{R}(t) = \mathcal{F}_t(t)/\mathcal{F}_0(t)$ is dominated by the rotation-term contribution $\mathcal{F}_{\gamma}$ in the long-time limit as 
\begin{equation}
    \mathcal{R}(t) = \frac{\mathcal{F}_t(t)}{\mathcal{F}_0(t)} 
    \sim \frac{F(t)^2}{t^2 f(t)},
\end{equation}
indicating that the achievable metrological enhancement is entirely determined by the modulation envelope $f(t)$. Specifically, a polynomial driving $f(t) \sim \mathcal{O}(t^n)$ delivers a genuine enhancement that scales as $\mathcal{R}(t) \sim \mathcal{O}(t^n)$, whereas an exponential modulation $f(t) \sim \mathcal{O}(e^t)$ yields a accelerated scaling of $\mathcal{R}(t) \sim \mathcal{O}(t^{-2}e^t)$.
This result indicates that by engineering the temporal modulation, one can reprogram arbitrary scaling laws governing measurement precision within energy and time constraints. To illustrate this capability, we subsequently analyze two specific regimes, linear [$f(t) = 1+at$] and exponential [$f(t) = e^{at}$] modulations, which manifest as linear and exponential enhancements in sensitivities, respectively.

\begin{figure}
    \centering
    \includegraphics[width=\linewidth]{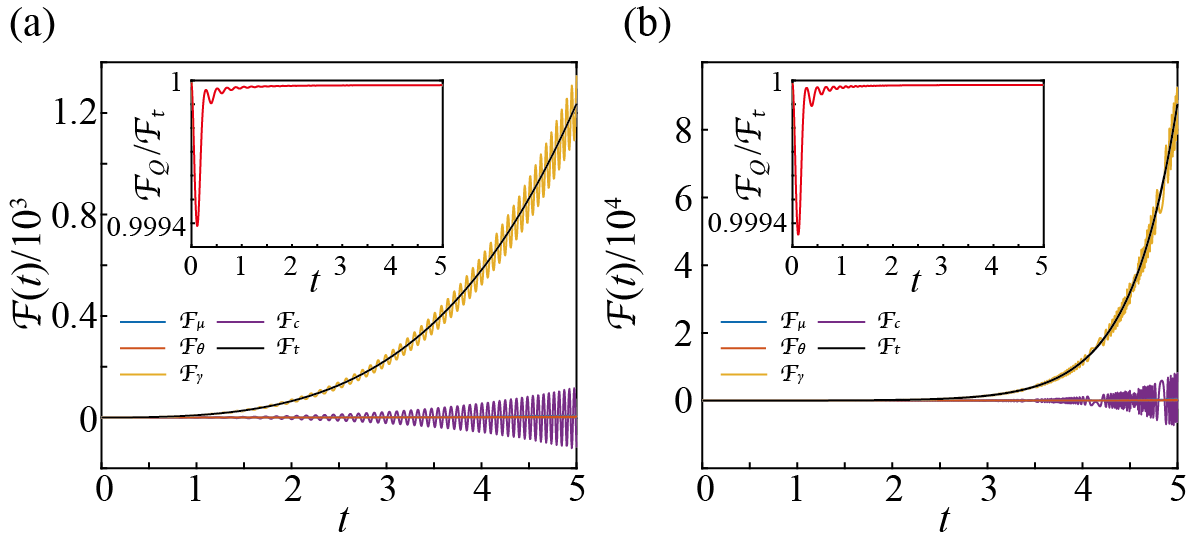}
    \caption{Time evolution of the total QFI $\mathcal{F}_t(t)$ (black lines) evaluated under the WKB approximation, decomposed into its constituent contributions: the shearing ($\mathcal{F}_\mu$), squeezing ($\mathcal{F}_\theta$), rotation ($\mathcal{F}_\gamma$), and cross-correlation ($\mathcal{F}_c$) terms. The oscillator temporal dynamics are driven by (a) a linear frequency modulation $f(t) = 1+at$ and (b) an exponential frequency modulation $f(t)=e^{at}$, with base frequency $\omega_0 = 10$ and modulation rate $a = 1$. Insets: Time dependence of the ratio $\mathcal{F}_Q/\mathcal{F}_t$, where $\mathcal{F}_Q$ is the CFI extracted from the optimal real-time homodyne measurement $\hat{Q}$.}
    \label{fig1}
\end{figure}

\textit{Example I: Linear Frequency Modulation.---}To concretely illustrate our framework, we examine a frequency estimation protocol utilizing a QHO subject to a linear frequency modulation, $\Omega(t)=\omega_0(1+at)$, where $a$ controls the modulation rate. In the adiabatic regime, the WKB approximation yields the explicit auxiliary function
\begin{equation}
    z(t)=\frac{1}{\sqrt{1+at}}
    \left[\left(1-\frac{ia}{4\omega_0}\right)e^{i\Phi(t)}
    +\frac{ia}{4\omega_0}e^{-i\Phi(t)}\right],
\end{equation}
where the accumulated dynamical phase is $\Phi(t) = \omega_0 F(t)$ with the integrated profile $F(t) = t + a t^2/2$. Substituting $z(t)$ into Eq.~(\ref{eq4}) completely determines the dynamical parameters and the resulting QFI. 
To rigorously validate this WKB analytical treatment, we compare it against exact numerical simulations of the dynamics; the exact QFI coincides with the WKB approximation~\cite{sm}.
As depicted in Fig.~\ref{fig1}(a), the rotational contribution $\mathcal{F}_{\gamma}$ overwhelmingly dominates the total information content, aligning with our asymptotic scaling predictions. 
Crucially, the final state $|\Psi\rangle_{\omega_0}$ remains strictly Gaussian, enabling efficient readout via quadrature detections that nearly saturate the QCRB. For a generalized quadrature observable $\hat{Q} = \hat{x} \cos q+\hat{p} \sin q$, the corresponding classical Fisher information (CFI) reads $\mathcal{F}_Q = \frac{(\partial_{\omega_0} \langle \hat{Q}\rangle)^2}{\Delta^2 \hat{Q}} + \frac{(\partial_{\omega_0} \Delta^2 \hat{Q})^2}{2 (\Delta^2 \hat{Q})^2}$. 
As shown in the inset of Fig.~\ref{fig1}(a), the ratio of the CFI to the total QFI rapidly approaches unity and remains above $0.999$, demonstrating that a straightforward homodyne measurement protocol can nearly saturate the QCRB.

\textit{Example II: Exponential Frequency Modulation.---}To access more aggressive scaling laws, we next consider an exponential modulation profile, $\Omega(t)=\omega_0 e^{at}$. Under the same adiabatic condition, $z(t)$ takes the form
\begin{equation}
    z(t) = e^{-at/2}\left[ (1-i\lambda/2)e^{i\Phi(t)}+i\lambda e^{-i\Phi(t)}/2  \right],
\end{equation}
with $\lambda = a/(2\omega_0)$, and the integrated profile $F(t) = \omega_0\left(e^{at}-1\right)/a$. Proceeding as before, we extract the total QFI. As corroborated by Fig.~\ref{fig1}(b), the rotational term $\mathcal{F}_{\gamma}$ again dictates the overarching information scaling, and the analytical WKB predictions remain in exact agreement with numerical simulations~\cite{sm}. 
Furthermore, this ultimate precision remains experimentally viable: the CFI-to-QFI ratio remains above $0.999$ [inset of Fig.~\ref{fig1}(b)], confirming that the precision limit can be robustly saturated via optimized homodyne measurements, irrespective of the specific modulation waveform.

\begin{figure}[t]
    \centering
    \includegraphics[width=\linewidth]{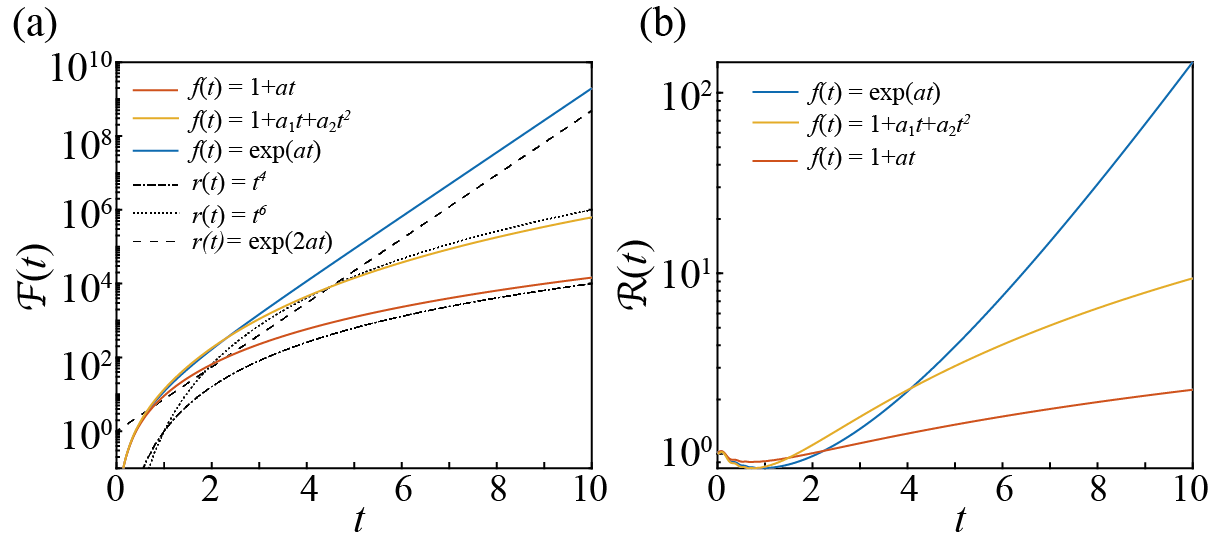}
    \caption{ Time evolution of (a) the QFI $\mathcal{F}_t(t)$ and (b) the QFI enhancement ratio $\mathcal{R}(t)$ for an initial coherent state with $\alpha=1$ and a base frequency $\omega_0 = 10$. Results are compared across three distinct modulation profiles: linear ($f(t) = 1+a t$, red lines), quadratic ($f(t) = 1+a_1 t + a_2 t^2$, yellow lines), and exponential ($f(t) = e^{at}$, blue lines). The black reference lines in (a) indicate the asymptotic scalings $t^4$, $t^6$, and $e^{2at}$. The modulation parameters are set to unity ($a = a_1 = a_2 = 1$).}
    \label{fig2}
\end{figure}

\textit{Comparison of temporal modulation strategies.---}To systematically benchmark the metrological capabilities of distinct temporal modulations, we comparatively analyze the QFI dynamics under linear [$f(t)=1+at$], quadratic [$f(t)=1+a_1t+a_2t^2$], and exponential [$f(t)=e^{at}$] frequency modulations. As depicted in Fig.~\ref{fig2}(a), the absolute precision scaling is directly governed by the temporal profile $F(t)^2$. Transitioning from polynomial to exponential modulation envelopes systematically enhances the asymptotic scaling of the QFI, thereby elevating the ultimate precision bounds.

To ensure that the observed advantage is not a trivial consequence of injecting additional energy into the oscillator, we evaluate the enhancement ratio $\mathcal{R}(t)$ in Fig.~\ref{fig2}(b), which quantifies the genuine metrological gain under strict energy and evolution-time constraints. Importantly, the hierarchical ordering of the modulation profiles remains robust against this energy penalty, where faster-growing driving envelopes consistently yield higher enhancement ratios. Most notably, the exponential profile maintains a pronounced advantage over its polynomial counterparts. This behavior confirms that the precision enhancement enabled by temporal control does not rely on an increased energy overhead. Instead, the time-dependent modulation unlocks a more efficient dynamical pathway for parameter encoding, optimizing metrological performance within resource-limited regimes.

\textit{Conclusion.---}In this Letter, we have demonstrated that arbitrary scaling enhancements in frequency estimation can be achieved by dynamically encoding an unknown base frequency into a QHO via temporal modulation. This approach modifies the mechanism of dynamical phase accumulation, providing a distinct pathway for optimizing metrological information acquisition. Specifically, for a temporal modulation profile $f(t)$ and its corresponding time-integral $F(t)$, the QFI scales as $\mathcal{O}(F(t)^2)$.
To establish a fair comparison with conventional time-independent encoding, we introduced the enhancement ratio $\mathcal{R}(t)$ to quantify the genuine metrological gain under strict energy and evolution-time constraints, yielding a genuine enhancement scaling as $\mathcal{R}(t) \sim \mathcal{O}(F(t)^2 / [f(t)t^2])$.
The universality of this framework was verified through a systematic analysis of linear, quadratic, and exponential modulation profiles. Furthermore, we demonstrated that these theoretical limits can be asymptotically saturated utilizing standard optimized homodyne detection. Ultimately, this work establishes that deterministically engineered precision scalings can be achieved by harnessing temporal modulation as a metrological resource without any additional energy overhead, thereby providing a versatile paradigm for optimizing continuous-variable quantum sensors and advancing high-efficiency quantum metrology across diverse experimental architectures.

A crucial distinction between our temporal modulation framework and quantum control approaches to metrology lies in what is being controlled. Control protocols usually enhance sensitivity by adding auxiliary Hamiltonian terms, engineering dissipation, or applying fast pulse sequences, all of which modify the generator of parameter encoding and often require prior knowledge of the parameter to design optimal controls~\cite{Yuan2015,Pang2017,Sekatski2017,Guo2019,Liu2024}. In contrast, our protocol directly modulates the parameter itself, with the modulation profile being fixed, deterministic, and completely independent of the unknown base frequency $\omega_0$. This has three key advantages: (i) no prior knowledge of $\omega_0$ is required to implement the optimal sensing strategy; (ii) the physical Hamiltonian structure remains unchanged, enabling a transparent resource accounting; and (iii) the protocol is open-loop and experimentally straightforward, requiring only a smoothly tunable oscillator frequency. Additionally, unlike fast-control or stroboscopic protocols that require rapid pulse sequences and often demand real-time feedback, our approach operates in the experimentally friendly adiabatic regime. The demonstrated scaling enhancements emerge purely from the cumulative phase imprinted by a slowly varying modulation, without requiring control bandwidth exceeding the system's natural frequency. This makes our scheme readily implementable across diverse continuous-variable platforms where smooth frequency sweeps are routine.

\textit{Acknowledgments.---}This work was supported by Quantum Science and Technology-National Science and Technology Major Project (Grant No. 2024ZD0302401 and No. 2021ZD0301500), National Natural Science Foundation of China (No. 12125402 and No. 12534016), and Beijing Natural Science Foundation (Grant No. Z240007).

\bibliography{ref_frequency_QHO}

@article{sm,
	journal = {See Supplemental Material for detailed derivations of the effective dynamics of a quantum harmonic oscillator with time-dependent frequency, the general framework of the temporal-modulation quantum metrology protocol, and the quantum Fisher information together with the corresponding enhancement ratio for linear, quadratic, and exponential frequency modulations}}

@article{Thew2007,
	author = {Gisin, Nicolas and Thew, Rob},
	date = {2007/03/01},
	doi = {10.1038/nphoton.2007.22},
	id = {Gisin2007},
	isbn = {1749-4893},
	journal = {Nat. Photon.},
	number = {3},
	pages = {165--171},
	title = {Quantum communication},
	url = {https://doi.org/10.1038/nphoton.2007.22},
	volume = {1},
	year = {2007}}

@article{chen2021,
  title={Review on quantum communication and quantum computation},
  author={Chen, Jiajun},
  journal={J. Phys.: Conf. Ser.},
  volume={1865},
  number={2},
  pages={022008},
  year={2021},
  publisher ={IOP Publishing},
  url = {https://iopscience.iop.org/article/10.1088/1742-6596/1865/2/022008/meta},
  doi = {10.1088/1742-6596/1865/2/022008}
}

@inbook{Aharonov1999,
author = {Dorit Aharonov},
title = {Quantum computation},
booktitle = {Annual Reviews of Computational Physics VI},
pages = {259-346},
doi = {10.1142/9789812815569_0007},
year = {1999},
URL = {https://www.worldscientific.com/doi/abs/10.1142/9789812815569_0007}
}

@article{Hansch2002,
	author = {Udem, Th. and Holzwarth, R. and H{\"a}nsch, T. W.},
	date = {2002/03/01},
	doi = {10.1038/416233a},
	id = {Udem2002},
	isbn = {1476-4687},
	journal = {Nature},
	number = {6877},
	pages = {233--237},
	title = {Optical frequency metrology},
	url = {https://doi.org/10.1038/416233a},
	volume = {416},
	year = {2002}}

@article{Rafal2016,
  title = {Ultimate Precision Limits for Noisy Frequency Estimation},
  author = {Smirne, Andrea and Ko\l{}ody\ifmmode \acute{n}\else \'{n}\fi{}ski, Jan and Huelga, Susana F. and Demkowicz-Dobrza\ifmmode \acute{n}\else \'{n}\fi{}ski, Rafa\l{}},
  journal = {Phys. Rev. Lett.},
  volume = {116},
  issue = {12},
  pages = {120801},
  numpages = {6},
  year = {2016},
  month = {Mar},
  publisher = {American Physical Society},
  doi = {10.1103/PhysRevLett.116.120801},
  url = {https://link.aps.org/doi/10.1103/PhysRevLett.116.120801}
}

@article{Cappellaro2017,
  title = {Quantum sensing},
  author = {Degen, C. L. and Reinhard, F. and Cappellaro, P.},
  journal = {Rev. Mod. Phys.},
  volume = {89},
  issue = {3},
  pages = {035002},
  numpages = {39},
  year = {2017},
  month = {Jul},
  publisher = {American Physical Society},
  doi = {10.1103/RevModPhys.89.035002},
  url = {https://link.aps.org/doi/10.1103/RevModPhys.89.035002}
}

@article{Sloyd2018,
	author = {Pirandola, S. and Bardhan, B. R. and Gehring, T. and Weedbrook, C. and Lloyd, S.},
	date = {2018/12/01},
	doi = {10.1038/s41566-018-0301-6},
	id = {Pirandola2018},
	isbn = {1749-4893},
	journal = {Nat. Photon.},
	number = {12},
	pages = {724--733},
	title = {Advances in photonic quantum sensing},
	url = {https://doi.org/10.1038/s41566-018-0301-6},
	volume = {12},
	year = {2018}}

@article{Silberhorn2018,
  title = {Quantum-Limited Time-Frequency Estimation through Mode-Selective Photon Measurement},
  author = {Donohue, J. M. and Ansari, V. and \ifmmode \check{R}\else \v{R}\fi{}eh\'a\ifmmode \check{c}\else \v{c}\fi{}ek, J. and Hradil, Z. and Stoklasa, B. and Pa\'ur, M. and S\'anchez-Soto, L. L. and Silberhorn, C.},
  journal = {Phys. Rev. Lett.},
  volume = {121},
  issue = {9},
  pages = {090501},
  numpages = {6},
  year = {2018},
  month = {Aug},
  publisher = {American Physical Society},
  doi = {10.1103/PhysRevLett.121.090501},
  url = {https://link.aps.org/doi/10.1103/PhysRevLett.121.090501}
}

@article{Felicetti2020,
  title = {Critical Quantum Metrology with a Finite-Component Quantum Phase Transition},
  author = {Garbe, Louis and Bina, Matteo and Keller, Arne and Paris, Matteo G. A. and Felicetti, Simone},
  journal = {Phys. Rev. Lett.},
  volume = {124},
  issue = {12},
  pages = {120504},
  numpages = {5},
  year = {2020},
  month = {Mar},
  publisher = {American Physical Society},
  doi = {10.1103/PhysRevLett.124.120504},
  url = {https://link.aps.org/doi/10.1103/PhysRevLett.124.120504}
}

@article{Gessner2025,
doi = {10.1088/1361-6633/ae00d8},
url = {https://doi.org/10.1088/1361-6633/ae00d8},
year = {2025},
month = {oct},
publisher = {IOP Publishing},
volume = {88},
number = {10},
pages = {106001},
author = {Fadel, Matteo and Roux, Noah and Gessner, Manuel},
title = {Quantum metrology with a continuous-variable system},
journal = {Rep. Prog. Phys.}
}

@article{Maccone2004,
author = {Vittorio Giovannetti  and Seth Lloyd  and Lorenzo Maccone},
title = {Quantum-Enhanced Measurements: Beating the Standard Quantum Limit},
journal = {Science},
volume = {306},
number = {5700},
pages = {1330-1336},
year = {2004},
doi = {10.1126/science.1104149},
URL = {https://www.science.org/doi/abs/10.1126/science.1104149}}

@article{Maccone2006,
  title = {Quantum Metrology},
  author = {Giovannetti, Vittorio and Lloyd, Seth and Maccone, Lorenzo},
  journal = {Phys. Rev. Lett.},
  volume = {96},
  issue = {1},
  pages = {010401},
  numpages = {4},
  year = {2006},
  month = {Jan},
  publisher = {American Physical Society},
  doi = {10.1103/PhysRevLett.96.010401},
  url = {https://link.aps.org/doi/10.1103/PhysRevLett.96.010401}
}

@article{Maccone2011,
	title = {Advances in quantum metrology},
	volume = {5},
	copyright = {http://www.springer.com/tdm},
	issn = {1749-4885},
	url = {https://www.nature.com/articles/nphoton.2011.35},
	doi = {10.1038/nphoton.2011.35},
	number = {4},
	urldate = {2024-07-17},
	journal = {Nat. Photon.},
	author = {Giovannetti, Vittorio and Lloyd, Seth and Maccone, Lorenzo},
	month = {Apr},
	year = {2011},
	pages = {222--229}
}

@article{Mitchell2011,
  title={Interaction-based quantum metrology showing scaling beyond the Heisenberg limit},
  author={Napolitano, Mario and Koschorreck, Marco and Dubost, Brice and Behbood, Naeimeh and Sewell, R. J and Mitchell, Morgan W},
  journal={Nature},
  volume={471},
  number={7339},
  pages={486--489},
  year={2011},
  publisher={Nature Publishing Group UK London},
  url = {https://www.nature.com/articles/nature09778}
}

@article{Zoller2024,
	title = {Essay: {Quantum} {Sensing} with {Atomic}, {Molecular}, and {Optical} {Platforms} for {Fundamental} {Physics}},
	volume = {132},
	issn = {0031-9007},
	url = {https://link.aps.org/doi/10.1103/PhysRevLett.132.190001},
	doi = {10.1103/PhysRevLett.132.190001},
	number = {19},
	urldate = {2024-09-19},
	journal = {Phys. Rev. Lett.},
	author = {Ye, Jun and Zoller, Peter},
	month = {May},
	year = {2024},
	pages = {190001}
}

@article{Zelevinsky2024,
	title = {Quantum sensing and metrology for fundamental physics with molecules},
	volume = {20},
	issn = {1745-2473},
	url = {https://www.nature.com/articles/s41567-024-02499-9},
	doi = {10.1038/s41567-024-02499-9},
	number = {5},
	urldate = {2024-07-17},
	journal = {Nat. Phys.},
	author = {DeMille, David and Hutzler, Nicholas R. and Rey, Ana Maria and Zelevinsky, Tanya},
	month = {may},
	year = {2024},
	pages = {741--749}
}

@article{Gao2025,
  title = {Realization of Versatile and Effective Quantum Metrology Using a Single Bosonic Mode},
  author = {Pan, Xiaozhou and Krisnanda, Tanjung and Duina, Andrea and Park, Kimin and Song, Pengtao and Fontaine, Clara Yun and Copetudo, Adrian and Filip, Radim and Gao, Yvonne Y.},
  journal = {PRX Quantum},
  volume = {6},
  issue = {1},
  pages = {010304},
  numpages = {15},
  year = {2025},
  month = {Jan},
  publisher = {American Physical Society},
  doi = {10.1103/PRXQuantum.6.010304},
  url = {https://link.aps.org/doi/10.1103/PRXQuantum.6.010304}
}

@article{Scarlino2025,
  title = {Criticality-Enhanced Quantum Sensing with a Parametric Superconducting Resonator},
  author = {Beaulieu, Guillaume and Minganti, Fabrizio and Frasca, Simone and Scigliuzzo, Marco and Felicetti, Simone and Di Candia, Roberto and Scarlino, Pasquale},
  journal = {PRX Quantum},
  volume = {6},
  issue = {2},
  pages = {020301},
  numpages = {16},
  year = {2025},
  month = {Apr},
  publisher = {American Physical Society},
  doi = {10.1103/PRXQuantum.6.020301},
  url = {https://link.aps.org/doi/10.1103/PRXQuantum.6.020301}
}

@article{Matteo2025a,
  title = {Enhanced Quantum Frequency Estimation by Nonlinear Scrambling},
  author = {Montenegro, Victor and Dornetti, Sara and Ferraro, Alessandro and Paris, Matteo G. A.},
  journal = {Phys. Rev. Lett.},
  volume = {135},
  issue = {3},
  pages = {030802},
  numpages = {8},
  year = {2025},
  month = {Jul},
  publisher = {American Physical Society},
  doi = {10.1103/39bt-37yl},
  url = {https://link.aps.org/doi/10.1103/39bt-37yl}
}

@article{Matteo2025b,
	author = {Cavazzoni, Simone and Teklu, Berihu and Paris, Matteo G. A.},
	date = {2025/11/10},
	doi = {10.1038/s41534-025-01112-y},
	id = {Cavazzoni2025},
	isbn = {2056-6387},
	journal = {npj Quantum Inf.},
	number = {1},
	pages = {174},
	title = {Frequency estimation by frequency jumps},
	url = {https://doi.org/10.1038/s41534-025-01112-y},
	volume = {11},
	year = {2025}}

@article{Braun2020,
  title = {Quantum parameter estimation of the frequency and damping of a harmonic oscillator},
  author = {Binder, Patrick and Braun, Daniel},
  journal = {Phys. Rev. A},
  volume = {102},
  issue = {1},
  pages = {012223},
  numpages = {13},
  year = {2020},
  month = {Jul},
  publisher = {American Physical Society},
  doi = {10.1103/PhysRevA.102.012223},
  url = {https://link.aps.org/doi/10.1103/PhysRevA.102.012223}
}

@article{Caves2011,
  title = {Fundamental Quantum Limit to Waveform Estimation},
  author = {Tsang, Mankei and Wiseman, Howard M. and Caves, Carlton M.},
  journal = {Phys. Rev. Lett.},
  volume = {106},
  issue = {9},
  pages = {090401},
  numpages = {4},
  year = {2011},
  month = {Mar},
  publisher = {American Physical Society},
  doi = {10.1103/PhysRevLett.106.090401},
  url = {https://link.aps.org/doi/10.1103/PhysRevLett.106.090401}
}

@article{Wilson2019,
	author = {McCormick, Katherine C. and Keller, Jonas and Burd, Shaun C. and Wineland, David J. and Wilson, Andrew C. and Leibfried, Dietrich},
	date = {2019/08/01},
	doi = {10.1038/s41586-019-1421-y},
	id = {McCormick2019},
	isbn = {1476-4687},
	journal = {Nature},
	number = {7767},
	pages = {86--90},
	title = {Quantum-enhanced sensing of a single-ion mechanical oscillator},
	url = {https://doi.org/10.1038/s41586-019-1421-y},
	volume = {572},
	year = {2019}}

@article{Biercuk2021,
  title = {Quantum Oscillator Noise Spectroscopy via Displaced Cat States},
  author = {Milne, Alistair R. and Hempel, Cornelius and Li, Li and Edmunds, Claire L. and Slatyer, Harry J. and Ball, Harrison and Hush, Michael R. and Biercuk, Michael J.},
  journal = {Phys. Rev. Lett.},
  volume = {126},
  issue = {25},
  pages = {250506},
  numpages = {6},
  year = {2021},
  month = {Jun},
  publisher = {American Physical Society},
  doi = {10.1103/PhysRevLett.126.250506},
  url = {https://link.aps.org/doi/10.1103/PhysRevLett.126.250506}
}

@article{Leibfried2021,
  title = {Quantum Harmonic Oscillator Spectrum Analyzers},
  author = {Keller, Jonas and Hou, Pan-Yu and McCormick, Katherine C. and Cole, Daniel C. and Erickson, Stephen D. and Wu, Jenny J. and Wilson, Andrew C. and Leibfried, Dietrich},
  journal = {Phys. Rev. Lett.},
  volume = {126},
  issue = {25},
  pages = {250507},
  numpages = {6},
  year = {2021},
  month = {Jun},
  publisher = {American Physical Society},
  doi = {10.1103/PhysRevLett.126.250507},
  url = {https://link.aps.org/doi/10.1103/PhysRevLett.126.250507}
}

@article{Smerzi2019,
	author = {Wolf, Fabian and Shi, Chunyan and Heip, Jan C. and Gessner, Manuel and Pezz{\`e}, Luca and Smerzi, Augusto and Schulte, Marius and Hammerer, Klemens and Schmidt, Piet O.},
	date = {2019/07/02},
	doi = {10.1038/s41467-019-10576-4},
	id = {Wolf2019},
	isbn = {2041-1723},
	journal = {Nat. Commun.},
	number = {1},
	pages = {2929},
	title = {Motional Fock states for quantum-enhanced amplitude and phase measurements with trapped ions},
	url = {https://doi.org/10.1038/s41467-019-10576-4},
	volume = {10},
	year = {2019}}

@article{Kippenberg2019,
	author = {Shomroni, Itay and Qiu, Liu and Malz, Daniel and Nunnenkamp, Andreas and Kippenberg, Tobias J.},
	date = {2019/05/07},
	doi = {10.1038/s41467-019-10024-3},
	id = {Shomroni2019},
	isbn = {2041-1723},
	journal = {Nat. Commun.},
	number = {1},
	pages = {2086},
	title = {Optical backaction-evading measurement of a mechanical oscillator},
	url = {https://doi.org/10.1038/s41467-019-10024-3},
	volume = {10},
	year = {2019}}

@article{Zaidi1988,
  title = {Squeezing and frequency jump of a harmonic oscillator},
  author = {Hong-Yi, Fan and Zaidi, H. R.},
  journal = {Phys. Rev. A},
  volume = {37},
  issue = {8},
  pages = {2985--2988},
  numpages = {0},
  year = {1988},
  month = {Apr},
  publisher = {American Physical Society},
  doi = {10.1103/PhysRevA.37.2985},
  url = {https://link.aps.org/doi/10.1103/PhysRevA.37.2985}
}

@article{Rhodoes1989,
  title = {Squeezing in harmonic oscillators with time-dependent frequencies},
  author = {Ma, Xin and Rhodes, William},
  journal = {Phys. Rev. A},
  volume = {39},
  issue = {4},
  pages = {1941--1947},
  numpages = {0},
  year = {1989},
  month = {Feb},
  publisher = {American Physical Society},
  doi = {10.1103/PhysRevA.39.1941},
  url = {https://link.aps.org/doi/10.1103/PhysRevA.39.1941}
}

@article{Rego2020,
	author = {Tibaduiza, D. M. and Pires, L. and Szilard, D. and Zarro, C. A. D. and Farina, C. and Rego, A. L. C.},
	date = {2020/10/01},
	doi = {10.1007/s13538-020-00770-x},
	id = {Tibaduiza2020},
	isbn = {1678-4448},
	journal = {Braz. J. Phys.},
	number = {5},
	pages = {634--646},
	title = {A Time-Dependent Harmonic Oscillator with Two Frequency Jumps: an Exact Algebraic Solution},
	url = {https://doi.org/10.1007/s13538-020-00770-x},
	volume = {50},
	year = {2020}}

@article{Seleznyova1995,
  title = {Unitary transformations for the time-dependent quantum oscillator},
  author = {Seleznyova, Alla N.},
  journal = {Phys. Rev. A},
  volume = {51},
  issue = {2},
  pages = {950--959},
  numpages = {0},
  year = {1995},
  month = {Feb},
  publisher = {American Physical Society},
  doi = {10.1103/PhysRevA.51.950},
  url = {https://link.aps.org/doi/10.1103/PhysRevA.51.950}
}

@article{Caves1994,
  title = {Statistical distance and the geometry of quantum states},
  author = {Braunstein, Samuel L. and Caves, Carlton M.},
  journal = {Phys. Rev. Lett.},
  volume = {72},
  issue = {22},
  pages = {3439--3443},
  numpages = {0},
  year = {1994},
  month = {May},
  publisher = {American Physical Society},
  doi = {10.1103/PhysRevLett.72.3439},
  url = {https://link.aps.org/doi/10.1103/PhysRevLett.72.3439}
}

@article{Augusto2009,
	title = {Entanglement, {Nonlinear} {Dynamics}, and the {Heisenberg} {Limit}},
	volume = {102},
	copyright = {http://link.aps.org/licenses/aps-default-license},
	issn = {0031-9007},
	url = {https://link.aps.org/doi/10.1103/PhysRevLett.102.100401},
	doi = {10.1103/PhysRevLett.102.100401},
	number = {10},
	urldate = {2025-02-28},
	journal = {Phys. Rev. Lett.},
	author = {Pezzé, Luca and Smerzi, Augusto},
	month = {Mar},
	year = {2009},
	pages = {100401},
}

@article{Kok2010,
	title = {General {Optimality} of the {Heisenberg} {Limit} for {Quantum} {Metrology}},
	volume = {105},
	copyright = {http://link.aps.org/licenses/aps-default-license},
	issn = {0031-9007},
	url = {https://link.aps.org/doi/10.1103/PhysRevLett.105.180402},
	doi = {10.1103/PhysRevLett.105.180402},
	number = {18},
	urldate = {2025-02-09},
	journal = {Phys. Rev. Lett.},
	author = {Zwierz, Marcin and Pérez-Delgado, Carlos A. and Kok, Pieter},
	month = {Oct},
	year = {2010},
	pages = {180402},
}

@article{Lan2021,
  title = {Rapid Quantum Squeezing by Jumping the Harmonic Oscillator Frequency},
  author = {Xin, Mingjie and Leong, Wui Seng and Chen, Zilong and Wang, Yu and Lan, Shau-Yu},
  journal = {Phys. Rev. Lett.},
  volume = {127},
  issue = {18},
  pages = {183602},
  numpages = {6},
  year = {2021},
  month = {Oct},
  publisher = {American Physical Society},
  doi = {10.1103/PhysRevLett.127.183602},
  url = {https://link.aps.org/doi/10.1103/PhysRevLett.127.183602}
}

@article{Sillanp2018,
	author = {Ockeloen-Korppi, C. F. and Damsk{\"a}gg, E. and Pirkkalainen, J. -M. and Asjad, M. and Clerk, A. A. and Massel, F. and Woolley, M. J. and Sillanp{\"a}{\"a}, M. A.},
	journal = {Nature},
	number = {7702},
	pages = {478--482},
	title = {Stabilized entanglement of massive mechanical oscillators},
	volume = {556},
	year = {2018},
    URL = {https://doi.org/10.1038/s41586-018-0038-x}}

@article{Clerk2020,
	author = {Clerk, A. A. and Lehnert, K. W. and Bertet, P. and Petta, J. R. and Nakamura, Y.},
	journal = {Nat. Phys.},
	number = {3},
	pages = {257--267},
	title = {Hybrid quantum systems with circuit quantum electrodynamics},
	volume = {16},
	year = {2020},
    URL = {https://doi.org/10.1038/s41567-020-0797-9}}

@article{Yuan2015,
  title = {Optimal Feedback Scheme and Universal Time Scaling for Hamiltonian Parameter Estimation},
  author = {Yuan, Haidong and Fung, Chi-Hang Fred},
  journal = {Phys. Rev. Lett.},
  volume = {115},
  issue = {11},
  pages = {110401},
  numpages = {7},
  year = {2015},
  month = {Sep},
  publisher = {American Physical Society},
  doi = {10.1103/PhysRevLett.115.110401},
  url = {https://link.aps.org/doi/10.1103/PhysRevLett.115.110401}
}

@article{Guo2019,
  title = {Control-Enhanced Sequential Scheme for General Quantum Parameter Estimation at the Heisenberg Limit},
  author = {Hou, Zhibo and Wang, Rui-Jia and Tang, Jun-Feng and Yuan, Haidong and Xiang, Guo-Yong and Li, Chuan-Feng and Guo, Guang-Can},
  journal = {Phys. Rev. Lett.},
  volume = {123},
  issue = {4},
  pages = {040501},
  numpages = {6},
  year = {2019},
  month = {Jul},
  publisher = {American Physical Society},
  doi = {10.1103/PhysRevLett.123.040501},
  url = {https://link.aps.org/doi/10.1103/PhysRevLett.123.040501}
}

@article{Pang2017,
	author = {Pang, Shengshi and Jordan, Andrew N.},
	date = {2017/03/09},
	doi = {10.1038/ncomms14695},
	id = {Pang2017},
	isbn = {2041-1723},
	journal = {Nat. Commun.},
	number = {1},
	pages = {14695},
	title = {Optimal adaptive control for quantum metrology with time-dependent Hamiltonians},
	url = {https://doi.org/10.1038/ncomms14695},
	volume = {8},
	year = {2017}}

@article{Liu2024,
  doi = {10.22331/q-2024-12-18-1571},
  url = {https://doi.org/10.22331/q-2024-12-18-1571},
  title = {Efficient tensor networks for control-enhanced quantum metrology},
  author = {Liu, Qiushi and Yang, Yuxiang},
  journal = {{Quantum}},
  issn = {2521-327X},
  publisher = {{Verein zur F{\"{o}}rderung des Open Access Publizierens in den Quantenwissenschaften}},
  volume = {8},
  pages = {1571},
  month = {dec},
  year = {2024}
}

@article{Sekatski2017,
  doi = {10.22331/q-2017-09-06-27},
  url = {https://doi.org/10.22331/q-2017-09-06-27},
  title = {Quantum metrology with full and fast quantum control},
  author = {Sekatski, Pavel and Skotiniotis, Michalis and Ko{\l{}}ody{\'{n}}ski, Janek and D{\"{u}}r, Wolfgang},
  journal = {{Quantum}},
  issn = {2521-327X},
  publisher = {{Verein zur F{\"{o}}rderung des Open Access Publizierens in den Quantenwissenschaften}},
  volume = {1},
  pages = {27},
  month = {sep},
  year = {2017}
}

\end{document}